\documentclass{emulateapj}
\RequirePackage{epstopdf}

\shorttitle{LBV Paper}
\shortauthors{Lau et al.}

\usepackage{graphicx}
\usepackage{amsmath}
\usepackage[section] {placeins}

\newcommand{\beq}{\begin{equation}}
\newcommand{\eeq}{\end{equation}}

\begin{document}

\submitted{Accepted to APJ March 5, 2014}

\title{NATURE VERSUS NURTURE: LUMINOUS BLUE VARIABLE NEBULAE IN AND NEAR MASSIVE STELLAR CLUSTERS AT THE GALACTIC CENTER}

\author{R. M. Lau\altaffilmark{1},
T. L. Herter\altaffilmark{1},
M. R. Morris\altaffilmark{2},
J. D. Adams\altaffilmark{1,3}
}

\altaffiltext{1}{Astronomy Department, 202 Space Sciences Building, Cornell University, Ithaca, NY 14853-6801, USA}
\altaffiltext{2}{Department of Physics and Astronomy, University of California, Los Angeles, 430 Portola Plaza, Los Angeles, CA 90095-1547, USA}
\altaffiltext{3}{SOFIA Science Center, Universities Space Research Association, NASA Ames Research Center, MS 232, Moffett Field, CA 94035, USA}

\begin{abstract} 

Three Luminous Blue Variables (LBVs) are located in and near the Quintuplet Cluster at the Galactic Center:  the Pistol star, G0.120-0.048, and qF362. We present imaging at 19, 25, 31, and 37 $\mu$m of the region containing these three LBVs, obtained with SOFIA using FORCAST. We argue that the Pistol and G0.120-0.048 are identical ``twins" that exhibit contrasting nebulae due to the external influence of their different environments. Our images reveal the asymmetric, compressed shell of hot dust surrounding the Pistol Star and provide the first detection of the thermal emission from the symmetric, hot dust envelope surrounding G0.120-0.048. However, no detection of hot dust associated with qF362 is made. Dust and gas composing the Pistol nebula are primarily heated and ionized by the nearby Quintuplet Cluster stars. The northern region of the Pistol nebula is decelerated due to the interaction with the high-velocity (2000 km/s) winds from adjacent Wolf-Rayet Carbon (WC) stars. From fits to the spectral energy distribution (SED) of the Pistol nebula with the DustEM code we determine that the Pistol nebula is composed of a distribution of very small, transiently-heated grains ($10-\sim35$ $\AA$) having a total dust mass of $0.03$ $M_\odot$, and that it exhibits a gradient of decreasing grain size from the south to the north due to differential sputtering by the winds from the WC stars. The total IR luminosity of the Pistol nebula is $5.2\times10^5\,L_\odot$. Dust in the G0.120-0.048 nebula is primarily heated by the central star; however, the nebular gas is ionized externally by the Arches Cluster. Unlike the Pistol nebula, the G0.120-0.048 nebula is freely expanding into the surrounding medium. A grain size distribution identical to that of the non-sputtered region of the Pistol nebula satisfies the constraints placed on the G0.120-0.048 nebula from DustEM model fits to its SED and implies a total dust mass of $0.021$ $M_\odot$. The total IR luminosity of the G0.120-0.048 nebula is $\sim10^5\,L_\odot$. From Paschen-$\alpha$ and 6 cm observations we determine a total gas mass of 9.3 $M_\odot$ and 6.2 $M_\odot$ for the Pistol and G0.120-0.048 nebulae, respectively. Given the independent dust and gas mass estimates we find that the Pistol and G0.120-0.048 nebulae exhibit similar gas-to-dust mass ratios of $310^{+77}_{-52}$ and $293^{+73}_{-101}$, respectively. Both nebulae share identical size scales ($\sim 0.7$ pc) which suggests that they have similar dynamical timescales of $\sim10^4$ yrs, assuming a shell expansion velocity of $v_\mathrm{exp}=60$ km/s.

\end{abstract}

\maketitle

\section{Introduction}
Stars classified as Luminous Blue Variables (LBVs) exist in a very brief ($<10^5$ yrs) and extreme evolutionary phase nearing the end of their lifetimes. Given the brevity of their lifetimes, LBVs are extremely rare and only $\sim10$ have been confirmed within the Milky Way (Clark et al. 2005).  Of the known Galactic LBVs, only a few have been found to be associated with their birth-cluster (Pasquali et al. 2006). Remarkably, three LBVs are located in the vicinity of the "Quintuplet" Cluster in the Galactic Center, a site of recent massive star formation (Mauerhan et al. 2010; Figer, McLean \& Morris 1999). The Quintuplet Cluster also hosts several Wolf-Rayet and dozens of O and B stars, all of which are believed to have formed coevally (Figer, McLean \& Morris 1999). Since LBVs are thought to be the intermediate link between massive O stars and Wolf-Rayet stars (Langer et al. 1994), studying these three LBVs in the context of the Quintuplet Cluster provides unique insight into the evolution of massive stars.

The three LBVs in the $3' \times 3'$ vicinity of the Quintuplet Cluster are qF362 (Figer, McLean \& Morris 1999; Geballe et al. 2000), the Pistol Star (Figer et al. 1998), and G0.120-0.048 (Mauerhan et al. 2010) (hereafter referred to as LBV3). Although these LBVs exhibit similar luminosities ($\sim10^6\,L_\odot$) and wind velocities ($\sim 100$ km/s) the nature of their outflows varies drastically. LBVs are typically surrounded by nebulae composed of material from their outflows (Nota et al. 1995); however, qF362 exhibits no observable nebular emission from gas or dust whereas the Pistol Star and LBV3 have surrounding shells of gas and dust (Yusef-Zadeh \& Morris 1987; Figer et al. 1999b; Mauerhan et al. 2010). The morphology and flux observed from the ionized gas emission in the nebulae surrounding the Pistol Star and LBV3 are notably dissimilar: the Pistol nebula appears compressed and asymmetric about the Pistol Star and shows strong emission from ionized gas at its northern edge while the LBV3 nebula is circularly symmetric about LBV3 and appears to have a uniform emission measure (Mauerhan et al. 2010). From Paschen-$\alpha$ observations,  Figer et al. (1999) argue that the preferential ionization of the Pistol nebula is due to the proximity of the hot O and B stars of the Quintuplet Cluster located to the north of the nebula, and that the Pistol Star itself contributes less than $8\%$ of the total ionizing flux to the nebula. The compression of the Pistol nebula at the northern edge and the northern displacement of the Pistol Star from the center of the nebula are attributed to the interaction with the winds of the Wolf-Rayet Carbon (WC) stars in the Quintuplet Cluster. Figer et al. (1999) also confirm from radial velocity observations of the Pistol nebula that it was formed from material in the outflows from the Pistol star.

LBV3 and its nebula are located approximately $\sim2.8'$ south-west of the Pistol Star and the Quintuplet Cluster. The LBV3 nebula was recently discovered by Mauerhan et al. (2010) from the HST/NICMOS Paschen-$\alpha$ survey of the Galactic Center (Dong et al. 2011; Wang et al. 2010). K-band spectroscopy of LBV3 revealed its spectral similarities to the Pistol Star and qF362 and confirmed its nature as a true LBV. The spherical symmetry and uniform ionization of the LBV3 nebula strongly suggests that it is composed of material from its outflow. Like the Pistol nebula, LBV3 does not produce enough Lyman-continuum photons to ionize the nebula and is likely externally ionized primarily by the hot stars in the Arches Cluster. Although the LBV3 and Pistol nebulae exhibit different morphologies and fluxes in Paschen-$\alpha$ emission their near-identical size scale (r$\sim200''$) and wind velocities indicate that their nebular ages are very similar ($\sim10^4$ yrs).

The thermal infrared dust emission from the Pistol and LBV3 nebulae provide important insight into the differences and similarities between the two sources. From observations with ISOCAM-CVF Moneti et al. (1999) first characterized the thermal mid-IR emission from the dust composing the Pistol nebula. The dust in the Pistol nebula appeared much more uniformly illuminated than the ionized gas; it more closely resembles a compressed sphere than the shape of a ``pistol." Spitzer/IRAC 8 $\mu$m observations of the Pistol nebula (Stolovy et al. 2006) revealed a morphology identical to that found by Moneti et al. (1999). In the case of LBV3, no detection was made of the nebular dust component with Spitzer/IRAC (and no observations were made with ISOCAM-CVF). This non-detection of the dust in the LBV3 nebula at 8 $\mu$m presents an interesting dichotomy in the dust properties of the two LBV nebulae in this region. Coupled with the absence of any form of nebular emission from qF362, the differences between the two LBV nebulae can be attributed to the impact of the influence of the hot Quintuplet Cluster stars and/or differences in the local ambient medium.

In this paper we present 19.7, 25.2, 31.5, and 37.1 $\mu$m observations tracing the warm dust emission of the Pistol nebula and the first detection of the warm dust emission of the LBV3 nebula taken by FORCAST aboard the Stratospheric Observatory for Infrared Astronomy (SOFIA). We compare and contrast the dust properties, morphology, and energetics of the nebulae and address the cause of the apparent differences between the three LBVs in this $3' \times 3'$ region of the Galactic Center.

\section{Observations and Data Reduction}
\subsection{FORCAST Imaging}

Observations were made using FORCAST (Herter et al. 2012) on the 2.5 m telescope aboard SOFIA. FORCAST is a  $256 \times 256$ pixel dual-channel, wide-field mid-infrared camera sensitive from $5 - 40~\mu\mathrm{m}$ with a plate scale of $0.768''$ per pixel and field of view of $3.4'\,\times\,3.2'$.
The two channels consist of a short wavelength camera (SWC) operating at $5 - 25~\mu\mathrm{m}$ and a long wavelength
camera (LWC) operating at $28 - 40~\mu\mathrm{m}$. An internal dichroic beam-splitter enables simultaneous observation from both long and short wavelength cameras. A series of bandpass filters is used to image at selected wavelengths.

SOFIA/FORCAST observations of the Pistol nebula were made on Basic Science Flight 64 on June 8, 2011 (altitude $\sim$ 43,000 ft.) at 19.7, 31.4, and 37.1 $\mu\mathrm{m}$. Follow-up images of both the Pistol and LBV3 nebulae were taken on the OC1-B Flight 110 on July 1, 2013 (altitude $\sim$ 39,000 ft.) at 19.7, 25.2, 31.4, and 37.1 $\mu\mathrm{m}$. Measurements at 19.7 and 31.4 $\mu$m, as well as 25.2 and 31.4 $\mu$m for Flight 110,  were observed simultaneously in dual-channel mode, while
the 37.1 $\mu$m observations were made in single-channel mode. Chopping and nodding were used to remove  the sky and telescope thermal backgrounds. An asymmetric chop pattern was used to place the source on the telescope
axis, which eliminates optical aberrations (coma) on the source. The chop throw was $7'$ at a frequency of $\sim 4$ Hz.
The off-source chop fields (regions of low mid-infrared Galactic emission) were selected from the Midcourse Space Experiment (MSX) $21~\mu$m image of the Galactic Center.
The source was dithered over the focal plane to allow removal of bad pixels and to mitigate response variations.
The integration time at each dither position was $\sim 20$ sec. The quality of the images was consistent with
near-diffraction-limited imaging at $19.7 - 37.1~\mu$m; the full width at half maximum (FWHM) of the point spread function (PSF) was
$3.2''$ at $19.7~\mu$m and $4.6''$ at 37.1 $\mu$m.

The acquired 100 sec integrated images were reduced and combined at each wavelength according to the pipeline steps described
in Herter et al. (2012). The calibration factors that were applied to the data numbers were the average calibration factors derived from calibration observations taken over the OC1-B Flight series, adjusted to those of a flat spectrum ($\nu F_\nu = \rm{constant}$) source. The $3\sigma$
uncertainty in the calibration factors is $\pm20\%$.

\subsection{ISOCAM-CVF Flux Discontinuity}

The Infrared Space Observatory Camera (ISOCAM) Circular Variable Filter (CVF) observations of the Pistol nebula cover wavelengths ranging from 8 - 17 $\mu$m. The observations from ISOCAM-CVF have a plate scale of $1.5''$ and a field of view of $48''\,\times\,48''$. There is a factor of $\sim2.5$ discontinuity between the flux from the nebula at $\leq17$ $\\mu$m and $\geq19.7$ $\mu$m as determined by ISOCAM-CVF and FORCAST, respectively. Spitzer/IRAC observations of the Pistol at 8 $\mu$m also exhibit a discrepancy in flux when compared to ISOCAM. The Spitzer/IRAC 8 $\mu$m flux of the nebula and the star are factor of $\sim1.5$ and $\sim2$ greater than the ISOCAM flux measurements, respectively; however, another star in the field, qF 76 (Figer et al. 1999b), exhibits a consistent flux between the two observations. The 2.38 - 40 $\mu$m ISOCAM Short-Wavelength Spectrometer (SWS) spectrum of the Pistol nebula is used as a reference for the expected 17/19.7 $\mu$m flux ratio. To normalize the ISOCAM and FORCAST fluxes with the ISO-SWS spectrum the ISOCAM fluxes are scaled up by a factor of 1.4. 

The consistent flux from qF 76 suggests that the discrepancy is unlikely due to issues with calibration. We attribute the flux discrepancy to variability in the Pistol Star luminosity over the 9 yr time interval between the ISOCAM-CVF and Spitzer/IRAC observations. Based on the dust heating models where the Pistol Star contributes to $\sim15\%$ of the heating of the nebula (Sec. 3.2.3), decreasing its luminosity by a factor of 2 will decrease the 8 $\mu$m flux by a factor of $\sim1.4$. This is consistent with the observed flux discrepancy between Spitzer/IRAC and ISOCAM-CVF.

\section{Results and Analysis}

\begin{figure*}[h]
	\centerline{\includegraphics[scale=.6]{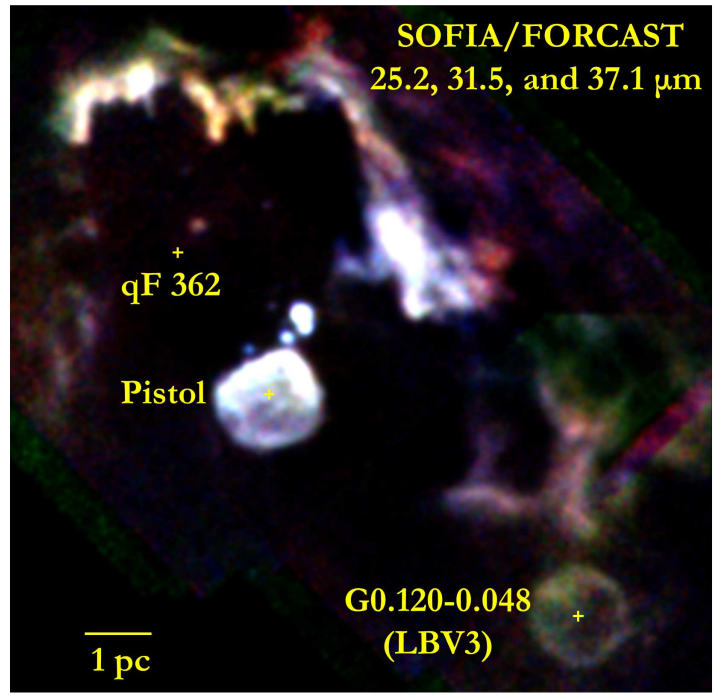}}
	\caption{25.2 (blue), 31.5 (green), and 37.1 (red) $\mu$m false color image of the Quintuplet Cluster containing the three luminous blue variables (LBVs) whose locations are marked by crosses.}
	\label{fig:FCimage}
\end{figure*}

The 19.7, 25.2, 31.5, and 37.1 $\mu$m images shown in Fig~\ref{fig:PistolLBV3Obs} reveal the warm dust emission from the Pistol nebula (a) and the LBV3 nebula (b). The equatorial coordinates (J2000) of the Pistol star and LBV3 are (266.563502, -28.834328) and (266.523436, -28.858866), respectively (Mauerhan et al. 2010). Fig.~\ref{fig:FCimage} shows the 25.2 (blue), 31.5 (green), and 37.1 (red) $\mu$m false color image of the Quintuplet Cluster region containing the three LBVs: qF362, the Pistol Star, and LBV3. We focus our analysis on the Pistol and LBV3 nebulae since we do not detect any obvious dust emission associated with qF362. The high signal-to-noise ratio of the Pistol nebula allows us to use a Richardson-Lucy deconvolution routine to improve the spatial resolution and provide a uniform 2.5'' full-width at half maximum (FWHM) Gaussian PSF at all wavelengths. For the deconvolution we derived the PSF to be Gaussian with the FWHM of the source GCS-4. In our analyses of the Pistol nebula we refer to the deconvolved images for the remainder of this paper. Given the lower signal-to-noise ratio of the LBV3 nebula we utilize the observed images in their native resolution in our analyses. 

In this section we present and analyze our results on the morphology and physical properties of dust contained in the Pistol and LBV3 nebulae. Their morphological and physical properties are summarized in Tab.~\ref{tab:PistolLBV3Prop}. For this paper, we assume that the Quintuplet Cluster is located in the Galactic center region at a distance of 8000 pc (Reid 1993).

\subsection{Interstellar Extinction}

Large column densities of dust and gas along lines of sight towards the Quintuplet Cluster at the Galactic Center lead to a large extinction ($A_V\sim30$) (Cardelli et al. 1989). In this paper we adopt the extinction curve constructed by Moneti et al. (2001, hereafter referred to as M2001) and utilize it at wavelengths longwards of 8 $\mu$m. An emissivity profile of the silicate features is used to define the M2001 curve between 8 and 24 $\mu$m. The silicate emissivity profile is a combination of the 9.7 feature derived by $\mu$ Cep observations (Roche \& Aitken 1984) and the 18 $\mu$m feature derived by the late type supergiant observations (Pegourie \& Papoular 1985) where the peak amplitude of the 9.7 $\mu$m optical depth is 2.9 and the ratio of the 9.7 and 18 $\mu$m optical depth amplitudes is $\tau_{18}/\tau_{9.7}=0.4$ (Simpson 1991). Longwards of 24 $\mu$m the M2001 curve follows a $\lambda^{-2}$ power law (Draine \& Lee 1984), which is typically assumed for Galactic Center extinction curves.

We additionally considered the Galactic Center extinction curve derived by Chiar \& Tielens (2006) from the Infrared Space Observatory (ISO) Short Wavelength Spectrometer (SWS) 2.38 - 40 $\mu$m spectra of GCS 3-I, one of the five Quintuplet Proper Members. Between 8 and 24 $\mu$m in their curve, Chiar \& Tielens (2006) adopt a silicate extinction profile with 9.7 and 18 $\mu$m strengths derived from fits to the spectrum of GCS 3-I assuming the underlying continuum emission can be approximated by a fourth-order polynomial. However, due to the large angular size of the ISO-SWS apertures ($14''\times20''$ and $14''\times27''$) the spectrum of GCS 3-I is contaminated with the prominent IR emission from the three other adjacent Quintuplet Proper Members. In their curve the extinction falls approximately as $\lambda^{-1}$ rather than the generally assumed $\lambda^{-2}$ power law.

\subsection{The Pistol Nebula}
\subsubsection{Pistol Morphology and Flux}
\begin{figure*}[c]
	\centerline{\includegraphics[scale=.3]{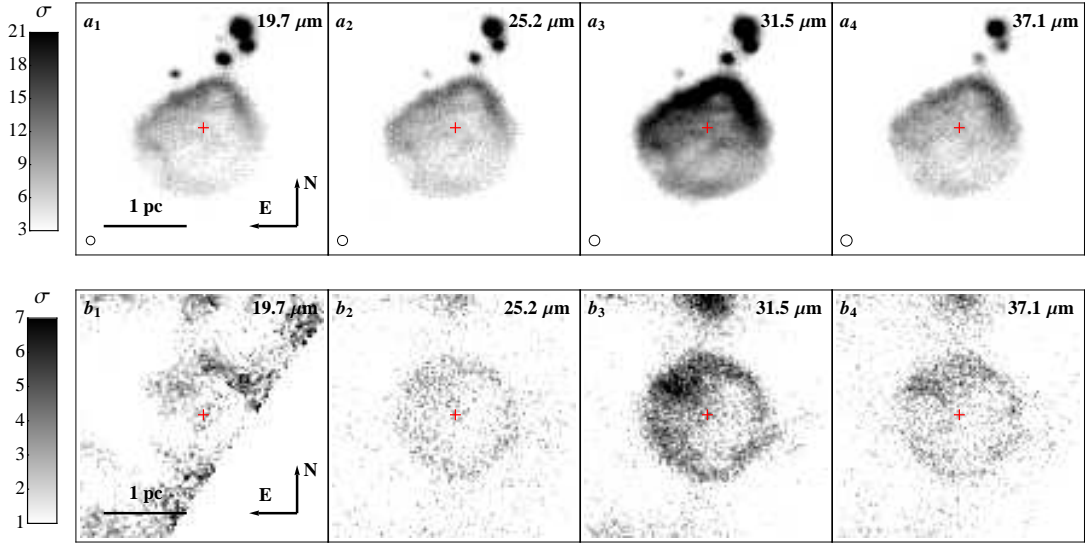}}
	\caption{Observed 19.7, 25.2, 31.5, and 37.1 $\mu$m images of the Pistol nebula (a) and the LBV3 nebula (b) with the positions of the respective stars marked by the red cross. The emission shown is above the 3-$\sigma$ flux for the Pistol nebula and above the 1-$\sigma$ flux for the LBV3 nebula. 3-$\sigma$ flux levels for the Pistol nebula images are 0.06, 0.09, 0.048, and 0.07 Jy/pixel at 19.7, 25.2, 31.5, and 37.1 $\mu$m, respectively; 1-$\sigma$ flux levels for the LBV3 nebula images are 0.007, 0.0275, 0.0172, and 0.032 Jy/pixel at 19.7, 25.2, 31.5, and 37.1 $\mu$m, respectively. The approximate beamsizes are shown in the lower left corner in each image of (a).}
	\label{fig:PistolLBV3Obs}
\end{figure*}

\begin{figure*}[h]
	\centerline{\includegraphics[scale=.5]{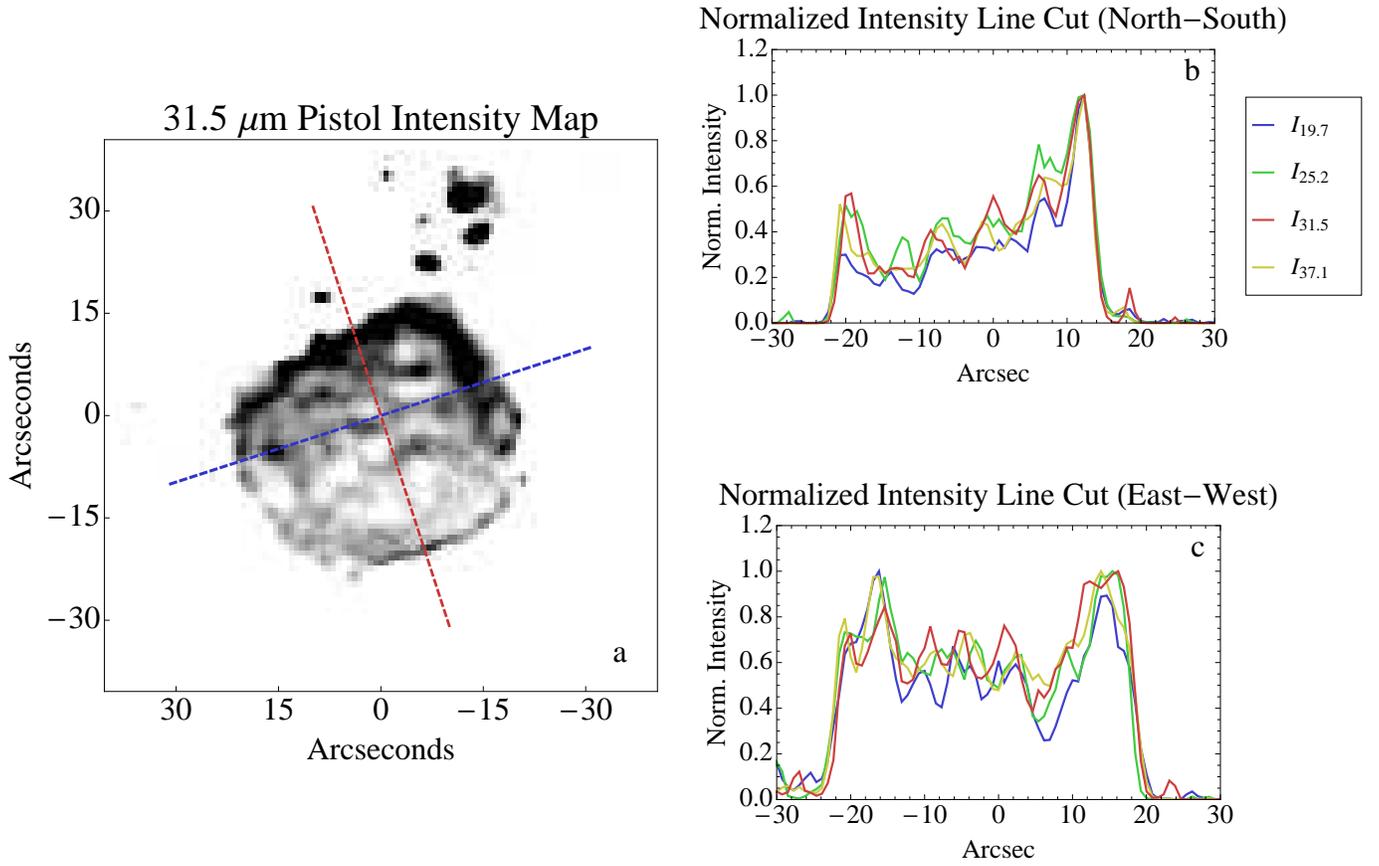}}
	\caption{(a) 31.5 $\mu$m image of the Pistol nebula overlaid with lines centered on the Pistol Star along which intensities of the 19.7, 25.2, 31.5, and 37.1 $\mu$m maps are extracted. (b) North-South and (c) East-West intensity profiles through the nebula normalized at the peak intensity.}
	\label{fig:PCuts}
\end{figure*}

Fig.~\ref{fig:PistolLBV3Obs}a shows that the location of the Pistol Star, which is indicated by the red cross, is slightly displaced to the northwest of the center of the nebula. The nebula itself appears compressed along the northern and western edges and has  East-West and North-South dimensions of 1.4 $\times$ 1.2 pc. The nearest distance in projection between the compressed northern and western edges of the nebula and the Pistol Star is $\sim0.4$ pc, whereas the projected distance to the eastern and southern edges is $\sim0.8$ pc. Dust emission from the nebula peaks along the northwest edge at all wavelengths; however, as can be seen in Fig.~\ref{fig:PistolLBV3Obs}a, the region of peak emission migrates from the north to the west with increasing wavelength. Unlike a typical HII region (Salgado et al. 2012), we find that the emission from the nebula appears nearly identical at each wavelength, which confirms its shell-like morphology. Normalized intensity line cuts through the nebula at each wavelength are shown in Fig.~\ref{fig:PCuts}. Fig.~\ref{fig:PCuts}b and c reveal the multi-wavelength similarities of the filaments and structures within the nebula. We also note that on the eastern edge of the nebula there is a faint ``ridge" of enhanced emission resembling a bow shock that possibly indicates an interaction with the winds of another nearby star.

Paschen-$\alpha$ contours (Dong et al. 2011) overlayed on the 31.5 $\mu$m intensity map (Fig.~\ref{fig:PIms}a) show that the ionized gas emission from the nebula closely traces the dust along the northern and western edges. The total Paschen-$\alpha$ flux from the nebula is $\sim 4.6\times10^{-10} \,ergs \,cm^{-2}\,s^{-1}$. We note that Paschen-$\alpha$ emission is more sensitive to density variations than dust emission since $I_{P\alpha}\propto n_e^2$ and $I_{Dust}\propto n_e$. As a first order correction to relate the two emission maps we ``square-root" the Paschen-$\alpha$ intensity. The ``square-root" Paschen-$\alpha$ intensity map closely traces the dust emission throughout the entire nebula which suggests the entire nebula is ionized. Since the Pistol Star has a luminosity and effective temperature of $3.3\times10^6\,L_\odot$ (Mauerhan et al. 2010) and $\sim12000$ K (Najarro et al. 2009) it will not be able to ionize the entire nebula; therefore, we require the ionizing flux from the nearby hot, Quintuplet Cluster stars to fully ionize the nebula (Figer et al. 1999).

\subsubsection{Observed and Predicted Pistol Dust Properties}

\begin{figure*}[h]
	\centerline{\includegraphics[scale=.4]{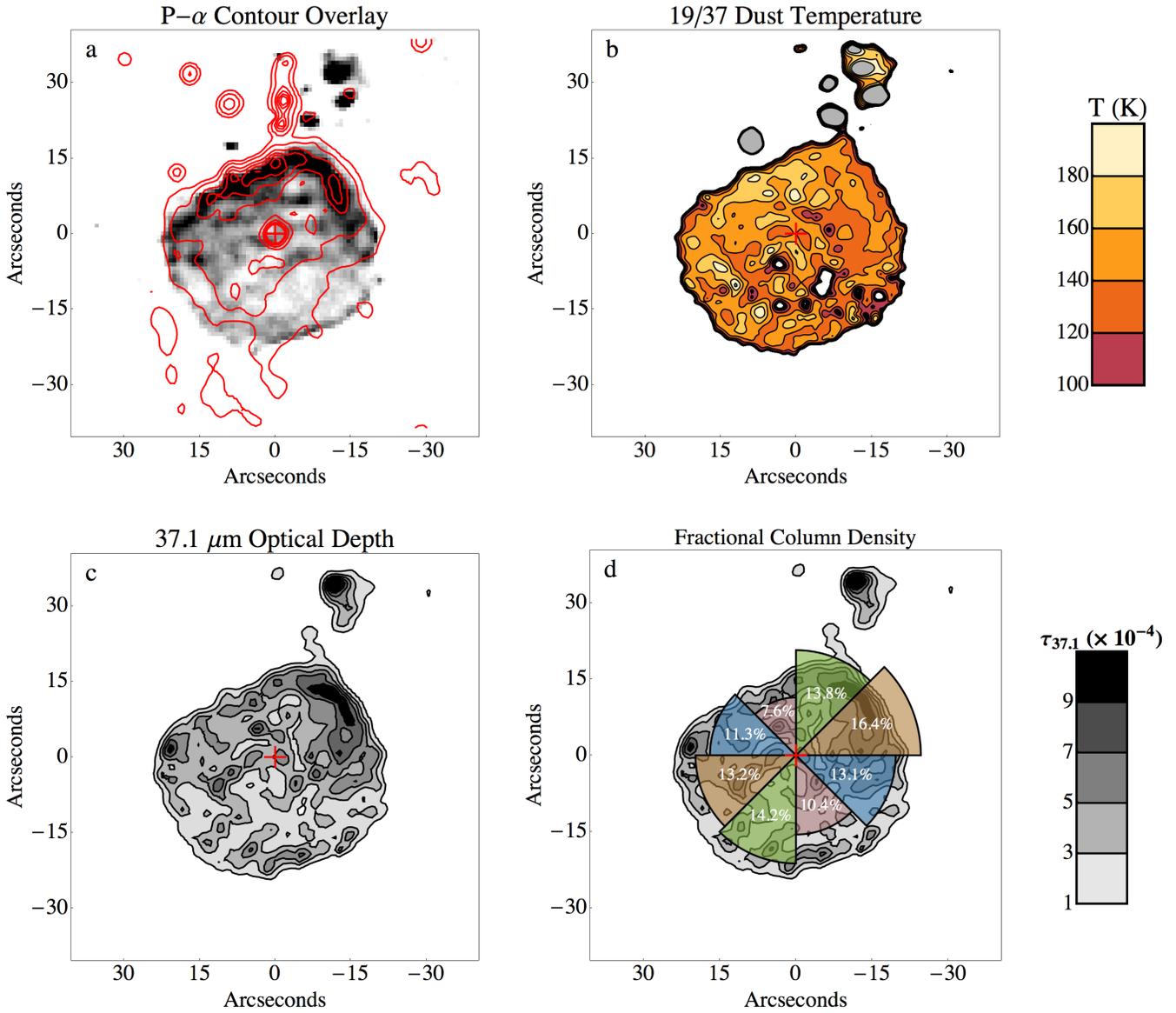}}
	\caption{ (a) 31.5  $\mu$m image of the Pistol nebula overlaid with Paschen-$\alpha$ contours with levels corresponding to 150, 250, 500, 750, 1000, and 1250 $\mu$Jy/pixel. (b) 19/37 dust temperature map of the Pistol nebula. (c) 37.1 $\mu$m optical depth map of the Pistol nebula. (d) same as (c) with the fractional column density overlaid. The length of the wedges correspond to the fractional integrated column density within the octant.}
	\label{fig:PIms}
\end{figure*}

We derive the 19.7/37.1 $\mu$m dust temperature map of the nebula (Fig.~\ref{fig:PIms}b) assuming that the emission is optically thin and can be approximated by a blackbody power-law modified by an emissivity of the form $\nu^\beta$. Since the Pistol Star is oxygen-rich, the dust in the nebula is likely composed primarily of silicate-type grains; therefore, we adopt a power-law index of $\beta=2$ which is consistent with the emissivity curve of silicates at wavelengths of $\lambda>19$ $\mu$m (Draine 2011). This leads to temperatures ranging from 100 - 180 K throughout the nebula, with an average of 150 $\pm10$ K, assuming a photometric error of $20\%$. The temperature peaks along filaments southeast of the projected location of the Pistol star. There is also a slight temperature gradient that gradually decreases from the northeast ($T_d\sim160 K$) edge to the southwest edge ($T_d \sim140 K$).

The optical depth at 37.1 $\mu$m for optically thin emission can be expressed as 

\beq
\tau_{37.1}= \frac{F_{37.1}}{\Omega_p\,B_\nu (T_d)},
\label{eq:OD}
\eeq
where $F_{37.1}$ is the 37.1 $\mu$m flux of the dust per pixel solid angle, $\Omega_p$, and $B_\nu (T_d)$ is the Planck function at dust temperature, $T_d$.  We produce a 37.1 $\mu$m optical depth map (Fig.~\ref{fig:PIms}c) using Eq.~\ref{eq:OD} and applying the values from the 19.7/37.1 $\mu$m temperature and the 37.1 $\mu$m intensity maps. The 37.1 $\mu$m optical depth peaks along the northwest edge of the nebula, with the largest value of $\tau_{37}\sim10^{-3}$, and decreases to the southeast to values of $\tau_{37}\sim3\times10^{-4}$. The average 37.1 $\mu$m optical depth throughout the nebula is $\sim5\times10^{-4}$. We find that the compressed edges of the nebula at the north and west are consistent with the regions of increased column density, which suggests that its morphology is influenced externally by winds from the nearby carbon-dominated Wolf-Rayet (WC) stars in the Quintuplet Cluster (Figer et al. 1999) and/or density gradients in the ambient medium. This claim is reinforced by the regions of low column density at the south and east of the nebula that are on the opposite side of the projected location of the Quintuplet Cluster. Fig.~\ref{fig:PIms}d shows the fractional column density integrated over the 8 octants about the Pistol Star and overlaid on the 37.1 $\mu$m optical depth map. We observe a slight north-south asymmetry in the column density in a sense that $\sim13\%$ more dust lies in the southern four octants of the nebula. 

Under the assumption that the dust in the nebula is solely heated by the Pistol Star and is in radiative equilibrium we can estimate a theoretical dust temperature, $T_d$, to compare with the observed temperatures. By balancing the power input from the Pistol Star and output by the dust grains we obtain the following expression for assumed spherical grains:

\beq
\pi  a^2Q_*(a)\, F_*=4 \pi  a^2\,Q_{\mathrm{dust}}(T_d,a)\,\sigma _{\mathrm{SB}}\,T_d{}^4,
\label{eq:Deq}
\eeq

where $Q_\mathrm{dust}$ is the dust emission efficiency averaged over the Planck function of dust grains of size $a$ and temperature $T_d$, $Q_*$ is the dust absorption efficiency averaged over the incident radiation field, and $F_*$ is the incident flux at the location of the dust. For $a=50\, \AA$-sized silicate-type grains and a 12000 K Kurucz model star with a log surface gravity of 4.5 and two times the solar metallicity (Kurucz 1993) as the heating source we have $Q_\mathrm{dust}(T_d, a)=6.5\times10^{-8} \left(\frac{a}{50\,\AA}\right)\,T_d^2$ and $Q_*(a)\sim0.04\left(\frac{a}{50\,\AA}\right)$ (Draine 2011). Assuming a distance of 0.7 pc from the star to the nebula and a total source luminosity of $3\,\times\,10^6\,\mathrm{L}_\odot$ (Mauerhan et al. 2010) we derive a theoretical equilibrium temperature of 88 K, which is significantly cooler than the observed 150 K. We note that since both $Q_\mathrm{dust}$ and $Q_*$ are approximately proportional to the grain size, $a$, the derived equilibrium temperature is not sensitive the grain size. The radiation from the star itself therefore does not provide enough energy to heat the dust in the nebula to the observed temperature.

When including the radiation from the Quintuplet Cluster stars, which we approximate as a 35000 K blackbody source with a total luminosity of $2\times 10^7\,L_\odot$ (Figer et al. 1999b) at a distance of 2.0 pc from the center of the nebula, we derive a dust equilibrium temperature of 130 K for the 50 $\AA$-sized silicate grains. The equilibrium temperature is slightly cooler than the observed temperature derived from the 19/37 flux ratio, which may be partially due to uncertainties in the stellar models; however, we argue that the discrepancy arises from the stochastic heating of very small grains, an effect that enhances the flux at shorter wavelengths. Our calculations therefore show that the observed dust temperatures are consistent with radiative heating by both the Pistol Star and Quintuplet Cluster stars, with the QC stars dominating the heating ($85\%$).

We find that the observed emission from the nebula at $\lambda<10$ $\mu$m is a factor of $\sim5$ greater than the flux predicted by a blackbody of temperature $T_\mathrm{dust} = 130$ K multiplied by the silicate dust emission efficiency $Q_\mathrm{Sil}(\lambda,a)$ (Draine \& Li 2001) fit to the observed FORCAST intensities (Tab.~\ref{tab:fluxes}). Again, we attribute the enhanced intensity at the shorter IR wavelengths to the presence of very small ($a<100\, \AA$), transiently heated grains in the nebula. Very small grains (VSGs) can reach temperatures much greater than equilibrium-heated large grains since they have small heat capacities that result in large temperature spikes after absorbing single photons. The emissivity of small grains can be characterized as

\beq
j_\nu=\int \mathrm{d}a \frac{\mathrm{d}n}{\mathrm{d}a} \int \mathrm{d}T \left(\frac{\mathrm{d}P}{\mathrm{d}T}\right)_a \sigma_\mathrm{abs}(\nu\mathrm{, }\,a)B_\nu(T),
\eeq

where $n$ is the dust density and $ \frac{\mathrm{d}P}{\mathrm{d}T}$ is a probability distribution function with $P(T)$ being the probability that a grain will have a temperature less than or equal to $T$. For large grains in radiative equilibrium the probability distribution function is simply a delta function at the equilibrium temperature. The steady state probability distribution function for small grains, which is much broader than the large grain function, can be solved for analytically (Guhathakurta \& Draine 1989; Draine \& Li 2001). 

\subsubsection{Pistol SEDs and IR Luminosity}

We utilize the DustEM (Compi{\`e}gne et al. 2011) code to model the SEDs from the northern, central, southern, and full regions of the nebula. The apertures used to extract the fluxes are overlayed on the 31.5 $\mu$m image of the nebula, as in Fig.~\ref{fig:PSEDfit}a. DustEM can model the emission from small, transiently heated grains and uses the formalism in Desert et al. (1986) to derive the temperature probability distribution function. In the models, we assume the dust is heated by both the Pistol Star and the hot stars in the Quintuplet Cluster, the latter of which also dominates the contribution of ionizing photons (Figer et al. 1999). 

Since the 3-dimensional spatial distribution of the Quintuplet Cluster stars with respect to the Pistol is not known we treat the cluster's radiation field in the DustEM models in two different ways: first, as arising from a point source-like distribution having a radiation field that declines as $r^{-2}$, and second, as an extended distribution that produces a plane-parallel field. The distance between the Quintuplet Cluster and the Pistol Star must be greater than the projected separation ($\sim1.5$ pc) since the back side of the nebula, the side furthest from the observer, has been decelerated more than the front side due to interactions with the winds from the nearby WC stars (Figer et al. 1999). We therefore adopt a factor of $\sqrt{2}$ to account for the distance projection, implying a 3D separation of $\sim2$ pc. For the plane-parallel field model we assume the entire nebula is heated by a radiation field identical to as the one expected at the center of the nebula in the $r^{-2}$ case ($d=2.0$ pc). Assuming the separation distance of $2.0$ pc we find that the Pistol Star contributes an average of $\sim15\%$ to heating the dust. The best DustEM model fits to each nebular region, assuming the plane-parallel field, are shown in Fig.~\ref{fig:PSEDfit}b - e with the minimum and maximum sizes of the small grains, $a_\mathrm{min,SG}$ and $a_\mathrm{max,SG}$, respectively. The $r^{-2}$ QC field case provides nearly identical results. The fitting parameters for both models are summarized in Tab.~\ref{tab:Pfits}.  

We find that the nebula is composed of a distribution of VSGs with sizes ranging from $10 - \sim35$ $\AA$. In addition to a gradient of decreasing flux from the north to the south of the nebula we also observe a decreasing $F_{8\mu m}/F_{25.2\mu m}$ ratio. The decreasing $F_{8\mu m}/F_{25.2\mu m}$ ratio in the plane parallel field case suggests the presence of increasingly larger grains towards the central and southern regions of the nebula. Since the total dust mass, $M_{Tot}$, is proportional to the dust flux we also observe a north-south gradient of decreasing mass in the plane-parallel radiation field case. A nearly identical north-south gradient is observed in the 37.1 $\mu$m optical depth map (Fig.~\ref{fig:PIms}c). We derive a total dust mass of $0.03$ $M_\odot$ and an integrated IR luminosity of $5.2\times10^5\,L_\odot$. 

We interpret the radiation field of the Quintuplet Cluster as plane-parallel since the Pistol Star is a member and the other cluster members are distributed within a region having an extent comparable to the distance separating them from the Pistol Star. The gradient of decreasing mass from the north to the south regions of the nebula from the models is also consistent with the observed 37.1 $\mu$m optical depth gradient. As mentioned previously, treating the Quintuplet Cluster as a point-like source with an $r^{-2}$ field does not result in significant quantitative and qualitative differences from the plane-parallel case. 

DustEM fits were attempted with the addition of an independent population of large, equilibrium-heated dust grains with $a=1000$ $\AA$. The fits indicate that in the central and southern regions of the nebula a population of large grains that comprises $\sim25\%$ of the total dust mass may be present. The peak flux contribution of the large grain population, however, is more than an order of magnitude smaller than that of the VSG population at the same wavelength. This suggests that even if large grains exist in the nebula, the total mass and IR luminosity are dominated by the VSGs. 

\subsection{The LBV3 Nebula}

\subsubsection{LBV3 Morphology and Flux}

\begin{figure*}[h]
	\centerline{\includegraphics[scale=.4]{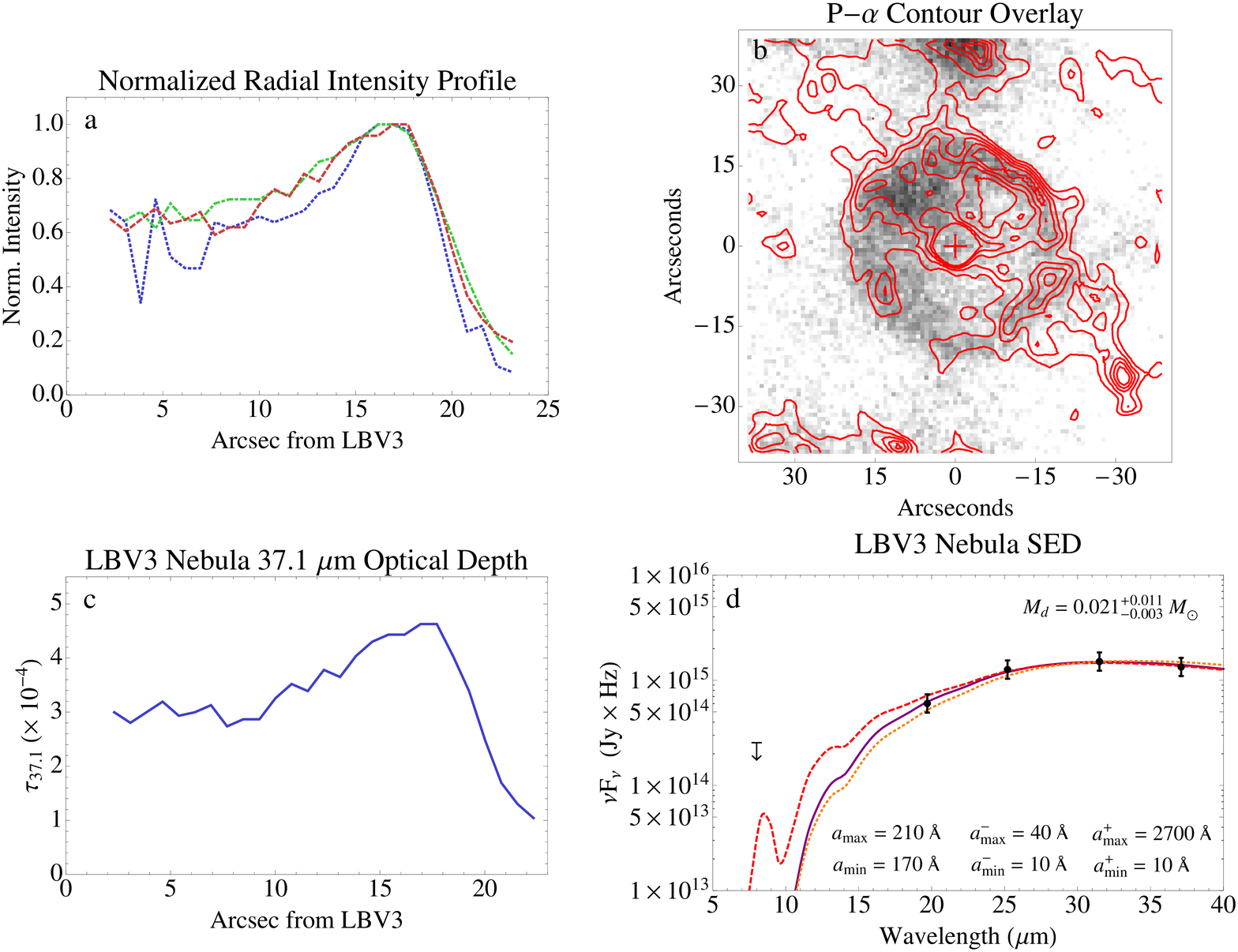}}
	\caption{(a) Normalized radial intensity profile of the LBV3 nebula at 25.2 (blue dotted), 31.5 (green dot-dashed), and 37.1 (red dashed) $\mu$m. (b) 31.5 $\mu$m image of the LBV3 nebula overlaid with Paschen-$\alpha$ contours with levels corresponding to 150, 175, 200, 225, and 250 $\mu$Jy/pixel. (c) Radial profile of the 37.1 $\mu$m optical depth of the LBV3 nebula. (d) DustEM model fits to the observed LBV3 FORCAST fluxes constraining the properties of the dust composing the LBV3 nebula.}
	\label{fig:LBVIms}
\end{figure*}

The dust emission from the LBV3 nebula shown in Fig.~\ref{fig:PistolLBV3Obs}b appears circularly symmetric about LBV3. Although the nebula exhibits a different morphology from the compressed, asymmetric Pistol nebula they share a similar size scale: the LBV3 nebula has a diameter of $\sim1.3$ pc. The consistent limb-brightening at all wavelengths indicates that the nebula has a shell-like morphology. Fig.~\ref{fig:LBVIms}a shows the normalized and azimuthally-averaged radial intensity profiles at 25.2, 31.5, and 37.1 $\mu$m centered on LBV3. The intensity throughout the nebula is not perfectly symmetric; Fig.~\ref{fig:PistolLBV3Obs}b shows that the intensity from the northeast quadrant is greater than the rest of the cavity by a factor of $\sim2$. The enhanced emission may be due to external heating contributions from other hot stars in the region or the asymmetric ejection of material from LBV3. Due to the low signal-to-noise ratio it is difficult to accurately map out the 2-dimensional temperature structure of the nebula to resolve this issue. Given the relative isolation of LBV3, we expect that LBV3 itself is the dominant source of heating of the dust in the nebula.

Paschen-$\alpha$ contours (Dong et al. 2011) overlayed on the 31.5 $\mu$m intensity map (Fig.~\ref{fig:LBVIms}b) reveal that the emission from the ionized gas closely traces the dust around the entire nebula, including the enhanced emission from the northeast quadrant. The dust and ionized gas are therefore likely confined to the same physical region within the nebula. The total Paschen-$\alpha$ flux from the nebula is $\sim 2\times10^{-10} \,ergs\,cm^{-2}\,s^{-1}$, which is a factor of $2.3$ less than that of the Pistol nebula. 

The uniformity of the Paschen-$\alpha$ emission and its strong similarities to that of the dust strongly suggest that the nebula is primarily ionized by a central stellar source. Since there are currently no direct radio observations of the nebula we scale the 6 cm flux of the Pistol nebula by the ratio of the Paschen-$\alpha$ fluxes from the nebulae to estimate the free-free emission at 6 cm from the LBV3 nebula.  Lang, Goss, \& Wood (1997) find the 6 cm emission from the Pistol nebula to be 0.5 Jy, which implies from our assumptions that the 6 cm emission is 0.22 Jy for the LBV3 nebula. 

The observed ratio of the Pistol nebula Paschen-$\alpha$ emission to that of the LBV3 nebula implies that $\sim2\times10^{48}$ Lyman-continuum photons $\mathrm{s}^{-1}$ are required to ionize the gas in the LBV3 nebula assuming it exhibits same electron temperature as that of the Pistol nebula, $T_e =$ 3600 K (Lang, Goss, \& Wood (1997). Adopting an electron temperature of 7000 K, a value similar to typical HII regions, only changes the Lyman-continuum flux estimate by less than $20\%$.

 With a luminosity and effective temperature of $4\times10^6 L_\odot$ (Mauerhan et al. 2010) and $12000$ K, respectively, LBV3 is unable to provide the ionizing flux to satisfy the observations. This suggests that the Arches Cluster, which produces $\sim4\times10^{51}$ (Figer et al. 2002) Lyman-continuum photons $\mathrm{s}^{-1}$ and is located $\sim 10$ pc away in projection from LBV3, is ionizing the nebula. Assuming the distance between LBV3 and the Arches Cluster is $\sqrt{2}\times10$ pc we derive that the cluster contributes $\sim2.5\times10^{48}$ Lyman-continuum photons $\mathrm{s}^{-1}$ to ionizing the nebula, which is consistent with required ionizing flux. Although the Quintuplet Cluster does not produce as many ionizing photons as the Arches it is closer in projection to the nebula and may also provide a significant fraction of the ionizing photons. Assuming a distance of $\sqrt{2}\times7$ pc between the QC and the nebula we find that it contributes $\sim10^{48}$ Lyman-continuum photons $\mathrm{s}^{-1}$. 

\subsubsection{Observed and Predicted LBV3 Dust Properties}

From the observed 19.7 and 37.1 $\mu$m fluxes of the nebula we derive a dust temperature of $105\pm8$ K assuming the silicate emissivity power law with an index of $\beta=2$. Given a dust temperature of 105 K and the 37.1 $\mu$m intensity profile we derive the radial 37.1 $\mu$m optical depth profile shown in Fig.~\ref{fig:LBVIms}c. We find that $\tau_{37}$ peaks at the edges of the nebula with a value of $\sim4.5\times10^{-4}$ and that it averages $\sim3.5\times10^{-4}$ over the entire nebula. The optical depth profile suggests that the dust is arranged in a shell-like morphology. 

We can estimate the equilibrium temperature expected for the dust in the nebula from Eq.~\ref{eq:Deq}, assuming that the heating is dominated by LBV3. For $0.005$ $\AA$-size silicate-type grains and a heating source with an effective temperature of 12000 K, a total luminosity of $4\times10^6 L_\odot$, and at a distance from the star of 0.7 pc, we derive a theoretical dust equilibrium temperature of 93 K. As with the Pistol nebula, the estimated dust equilibrium temperature may be slightly cooler than the observed temperature derived from the 19/37 flux ratio due to the presence of very small grains. 

Including the heating from the Arches Cluster, which is approximated as a 35000 K blackbody with a luminosity of $6.5\,\times\,10^{7}$ $L_\odot$ (Figer et al. 2002) at a presumed distance of $14$ pc from the nebula, we find that the dust equilibrium temperature rises to 98 K. This dust temperature estimate is more consistent with the observed temperature than in the case with only central heating by LBV3, which suggests that, while the dust heating is dominated by LBV3 ($70\%$), there is an additional radiative heating contribution from the Arches Cluster. This scenario is consistent with the nebula being externally ionized by the Arches Cluster since the LBV3 star has an effective temperature of 12000 K and is unable to provide enough ionizing photons.

\subsubsection{LBV3 SED and IR Luminosity}

We utilize the DustEM code to model the observed SED of the dust emission from the nebula and assume the nebula is symmetric about LBV3 and heated by both LBV3 and the Arches Cluster. LBV3 and the Arches Cluster are approximated as the same 12000 K Kurucz model star as for the Pistol model and a $6.5\times10^7\,L_\odot$ blackbody source (Figer et al. 2002) with an effective temperature of 35000 K, respectively. Fig.~\ref{fig:LBVIms}d shows three DustEM fits to the emission from the nebula at 19.7, 25.2, 31.5, and 37.1 $\mu$m: the best fit, the 1-$\sigma$ maximum grain size cutoff upper limit fit, $a^+$, and the 1-$\sigma$ maximum grain size cutoff lower limit fit, $a^-$. The 1-$\sigma$ upper limit of the minimum grain size cutoff is 600 $\AA$. Also shown in \ref{fig:LBVIms}d is the background continuum flux at 8 $\mu$m detected by Spitzer/IRAC. Owing to the non-detection of this background in the mid-IR, and to the ``flatness" of the LBV3 SED, we are unable to tightly constrain the grain size distribution of the nebula. The dust mass and integrated IR luminosity derived from the best-fit DustEM model are $\sim0.021^{+0.011}_{-0.003}$ $M_\odot$ and  $8_{-0.5}^{+1.5} \times10^4$ $L_\odot$, respectively. The fitting parameters and results of the model are summarized in Tab.~\ref{tab:Pfits}.

\subsection{Gas to Dust Mass Ratio}

We derive the total gas mass of the nebulae from the observed Paschen-$\alpha$ flux. For the Pistol nebula we approximate its volume as that of a spherical shell with a thickness of 0.1 pc as determined by the high-resolution Paschen-$\alpha$ observations by Figer et al. (1999) and an inner radius of 0.6 pc. Assuming that the entire Pistol nebula is ionized and exhibits a 6 cm flux of 0.5 Jy (Lang, Goss \& Wood 1997) we determine a total gas mass of 9.3 $M_\odot$. Figer et al. (1999) derive a slightly greater gas mass (11 $M_\odot$); however, they assume a smaller volume for the nebula and that only half the nebula is ionized. From our gas mass calculation and dust model fit to the SED ($\sim0.03$ $M_\odot$) we derive a gas-to-dust mass ratio of  $310^{+77}_{-52}$ for the Pistol nebula.

For the LBV3 nebula we adopt the same volume as the Pistol and estimate a total gas mass of $6.2$ $M_\odot$. Taking the ratio of the total gas mass with the dust mass determined from DustEM model fits to the LBV3 nebula SED ($\sim0.02$ $M_\odot$) provides a gas-to-dust ratio of $293^{+73}_{-101}$, which is similar to that of the Pistol.

\section{Discussion}

\subsection{LBV3 Symmetric Shell Intensity Model}
\begin{figure*}[h]
	\centerline{\includegraphics[scale=.6]{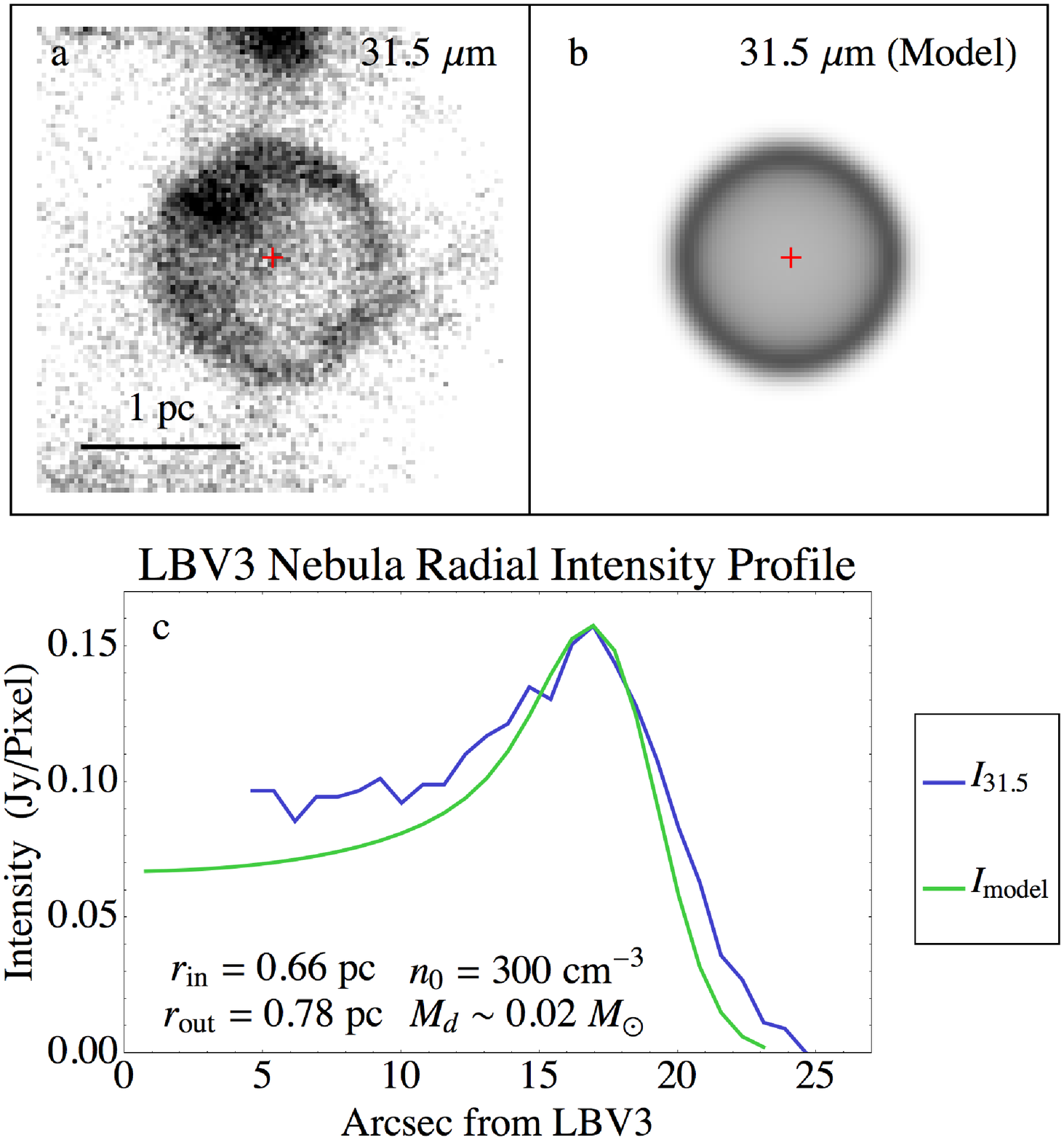}}
	\caption{ (a) Observed 31.5 $\mu$m image of the LBV3 nebula. (b) 31.5 $\mu$m symmetric shell intensity model. (c) Azimuthally averaged radial intensity profile of the LBV3 nebula extracted from the observed and modeled 31.5 $\mu$m images; the northeast quadrant of the nebula has been ignored in determining the observed radial profile.}
	\label{fig:LBV3Modcomp}
\end{figure*}

To understand the structure and heating of the LBV3 nebula we model the emission as originating from a spherically symmetric shell in which the dust is centrally heated by LBV3 and uniformly heated by the Arches Cluster. Input parameters are the inner shell radius, $r_1$, outer shell radius, $r_2$, the inner radius temperature, $T_0$,  the inner shell gas density, n$_0$, the exponent of the power-law radial density profile, $\alpha$, and the minimum and maximum grain size cutoffs. The gas-to-dust mass ratio is set to 293 and the minimum and maximum grain size cutoffs are identical to that of the southern Pistol nebula ($a_\mathrm{min}=10$ $\AA$ and $a_\mathrm{max}=60$ $\AA$). The remaining parameters are fit with the values shown in Tab.~\ref{tab:LBV3tab}. Fig.~\ref{fig:LBV3Modcomp}a and b show the 31.5 $\mu$m image of the LBV3 nebula and the intensity model convolved to the observed 31.5 $\mu$m PSF, respectively. We plot the azimuthally-averaged intensity profile of the model and the observed nebula excluding the northeast quadrant of enhanced emission in Fig.~\ref{fig:LBV3Modcomp}c. The slight discrepancy between the model and the observed intensities in the inner regions of the nebula ($<10''$) may be attributed to confusion with the emission from the lower part of the Sickle at the northern and southeastern regions of the nebula. The best-fit model predicts a gas density of 300 $cm^{-3}$ with a flat radial density profile between the inner and outer radii and a total dust mass of 0.02 $M_\odot$, consistent with the DustEM model fits to the observed SED. We note the degeneracy between the radial density power law and inner radius parameters: a model with a steep $\alpha=10$ radial density power-law and an inner shell radius near the center also provides an appropriate fit. Despite the degeneracy, the agreement between both $\alpha=0$ and $\alpha=10$ models  and the observed results show that the morphology of LBV3 is that of a shell where the density rises rapidly at some radius.

\subsection{Evidence for Complete Ionization of the Nebulae}

The similar morphologies of the dust distribution, which is fully illuminated in both nebulae, and the Paschen-$\alpha$ emission strongly suggests that the two nebulae are completely ionized. The absence of molecular material associated with the Pistol nebula (Serabyn \& Gusten 1991) also supports the claim of full ionization. From the Stromgren equation (Osterbrock \& Ferland 2006), we can estimate the depth that Lyman-continuum photons from the Quintuplet Cluster stars can penetrate into the Pistol nebula, $\Delta s$, assuming the stars produce $10^{51}$ ionizing photons per second (Figer et al. 1999) and are located $d_N= 2.0$ pc from the Pistol.

\beq
\Delta s=\left( \frac{N_\mathrm{LyC}}{4/3 \pi \alpha_B n_e^2}+d_N^3\right)^{1/3}-d_N
\label{eq:SE}
\eeq

Given a density of $n_e\sim700\,cm^{-3}$ for the Pistol nebula, which we derive from the 6 cm flux (Lang, Goss, \& Wood 1997) and the volume of the nebula, we find that the ionizing photons can penetrate to a depth of $\sim1.5$ pc which is roughly the diameter of the Pistol nebula and an order of magnitude greater than the observed thickness. Even if the nebula were located 5 pc away from the Quintuplet Cluster stars the Lyman-continuum photons would still be able to fully ionize the nebula. We therefore conclude that the Pistol nebula must be entirely ionized. 

For the LBV3 nebula we determine an ionization depth of $\sim 1.2$ pc assuming a density of $n_e\sim300\,cm^{-3}$ and that the Arches Cluster produces $4\times10^{51}$ ionizing photons per second. Despite the fact that the dust in the LBV3 nebula is primarily heated by the central star, the nebula is fully ionized externally assuming that the Lyman-continuum flux originates from the Arches Cluster as we have argued in Sec 3.3.1. 

\subsection{Evolution of the LBV Nebulae}

Both the Pistol and LBV3 nebulae exhibit similar sizes ($r\sim0.7$ pc), indicating that they also share identical dynamical timescales. An expansion velocity of 60 km/s (Figer et al. 1999), which is not unlike those found for other LBV nebulae (Kochanek 2011), gives a dynamical timescale of $\sim10^4$ yrs.

We interpret the nebulae as being primarily composed of dust and gas formed in the outflows of the stars rather than being swept-up interstellar material. As suggested by Lamers et al. (2001), the medium surrounding LBVs has been evacuated by the winds during their main-sequence phase and therefore exhibits densities of $\sim10^{-3}$ $\mathrm{cm}^{-3}$. This implies that for a spherical nebula with a radius of 0.7 pc a total of only $3.5\times10^{-5}\,M_\odot$ of material will be swept up, which is significantly less than their estimated total mass ($\sim6 -9\,M_\odot$). Additionally, in the case of the Pistol nebula, Lang et al. (1997) find an enhancement of helium with respect to the nearby Sickle HII region which suggests a stellar origin. The 

In the following subsections we address the distinguishing characteristics of all three LBVs in the context of their evolution and environment. 

\subsubsection{Dust Formation in LBV Outbursts}

Dust production around massive stars poses a difficult challenge to explain, given their greater luminosities and effective temperatures. We suggest that the formation scenario of the dust in the LBV nebulae is that proposed by Kochanek (2011): dust is formed during a short outburst phase ($\sim1000$ yrs) of the LBV with extreme mass-loss rates ($\sim 10^{-3}\,M_\odot$) and dense winds to shield dust formation regions from UV photons that would otherwise photoevaporate the small grains. In their quiescent state LBVs have too high an effective temperature ($\sim15000$ K) and too low a mass loss rate ($\sim10^{-5}\,M_\odot/yr$) for dust grains to form or survive. LBVs may also exist in cool states where they expand and exhibit effective temperatures of $\sim7000$ K; however, the soft UV photons will still photoevaporate small grains and thus impede dust formation. We therefore require a high mass-loss rate to both initiate the collisional growth of the dust grains as well as produce a dense wind to shield the grains. From the momentum conservation of the radiatively-driven winds, $\dot{M}v_\infty\sim\tau_V L/c$, we see that for $L=10^6\,L_\odot$ and $v_\infty=100$ km/s $\dot{M}$ must be $\gtrsim 10^{-3.5}\,M_\odot$ for $\tau_V \gtrsim 1$. Since the swept-up mass is negligible, the nebulae are therefore a reflection of the history of the mass-loss rate from the LBVs.

\subsubsection{Maximum Grain Size and Mass Loss Rate}

From the maximum grain size it is possible to approximate the mass-loss rate during grain formation, assuming the grains grow by accretion of gas particles with mass $m_g$ (Kochanek 2011).  Assuming $100\%$ sticking probability, the collisional growth rate can be expressed as

\beq
\frac{d N}{d t}=\frac{\dot{M}X_g \,\pi  a^2 \,v_t}{4 \pi  R^2 \,v_w\,m_g},
\label{eq:MD1}
\eeq

where $\dot{M}$ is the mass-loss rate, $X_g$ is the mass fraction of the condensible dust species, $v_t$ is the thermal velocity of the particles, and $v_w$ is the velocity of the stellar winds. The number of collisions with the gas particles can be related to the grain size by $N=\frac{\frac{4}{3}\pi a^3\rho _b}{m_g}$, where $\rho_b$ is the bulk density of a grain ($\sim$3 gm $\mathrm{cm}^{-3}$), and we can then derive the grain growth rate as

\beq
\frac{d a}{d t}=\frac{\dot{M}X_g\,v_t}{16 \pi  R^2\,v_w\,\rho _b}.
\label{eq:MD2}
\eeq

Since the gas will cool as it expands, $v_t$ can be expressed in terms of $R$, the radius, and $R_f$, the grain formation radius ($\sim10^{15}$ pc), as $v_t(R)=v_{t0} \left(\frac{R}{R_f}\right)^{-n}$, where $n=2/3$ for cooling by adiabatic expansion (Kochanek 2011). After integrating, the grain size as a function of radius can be expressed as

\beq
a(R)\approx \frac{\dot{M}X_g \,v_{t0}}{16 \pi  R_fv_w^2\rho _b\,(1+2/3)}\left[1-\left(\frac{R_f}{R}\right)^{1+2/3}\right].
\label{eq:MD3}
\eeq

Assuming that the grains stop growing and reach their maximum size when the growth rate drops to $10\%$ of the initial rate, or at a distance of $R\simeq 3R_f$, we find that

\begin{multline}
a_\mathrm{max}\sim 60\,\left(\frac{\dot{M}}{10^{-3}\,M_\odot/yr}\right)\,\left(\frac{X_g}{0.003}\right) \\  \left(\frac{v_{t0}}{1 \, \mathrm{km/s}}\right)\,\left(\frac{100 \,\mathrm{km/s}}{v_w}\right)^2\,\AA.
\label{eq:MD4}
\end{multline}

\subsubsection{Dust Evolution in the Pistol Nebula}

\begin{figure*}[h]
	\centerline{\includegraphics[scale=.4]{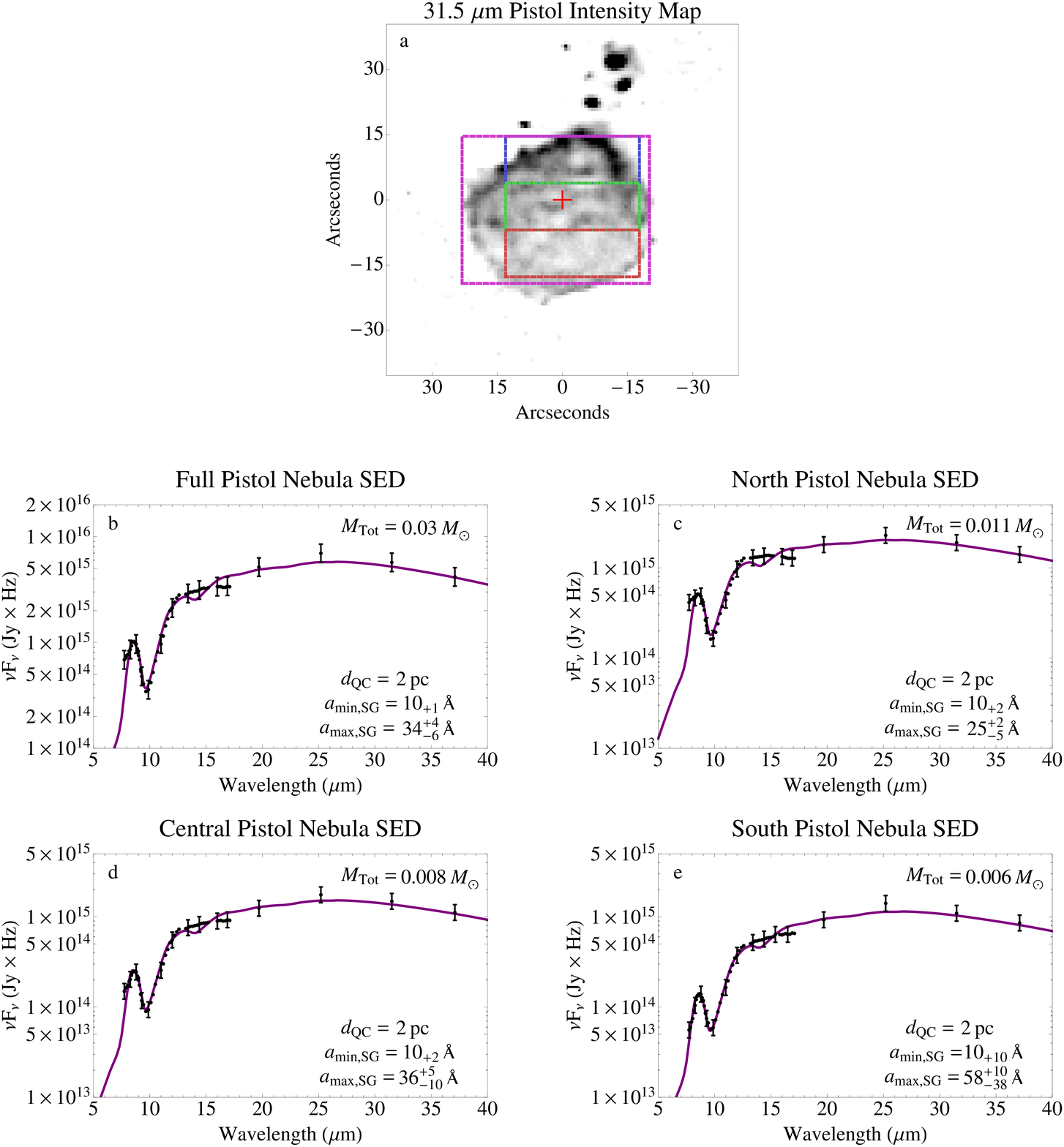}}
	\caption{(a) 31.5 $\mu$m image of the Pistol nebula overlaid with the north, central, south, and full apertures used to extract the fluxes. (b) - (e) Plane-parallel radiation field DustEM fits to the observed ISOCAM-CVF (7 - 17 $\mu$m) and FORCAST (19.7 - 37.1 $\mu$m) fluxes in units of Jy $\times$ Hz. $M_\mathrm{Tot}$, $d_\mathrm{QC}$, $a_\mathrm{min,SG}$, and $a_\mathrm{max,SG}$ are the total dust mass, distance to Quintuplet Cluster, minimum grain size, and maximum grain size, respectively. The errors in the grain size correspond to a 3-$\sigma$ deviation from the $\chi^2_\mathrm{min}$.}
	\label{fig:PSEDfit}
\end{figure*}

We suggest that the Pistol star produced silicate grains with sizes ranging from $10 - 60$ $\AA$ during the outburst that formed the nebula. The selection of this size distribution is based on the parameters of the SED fit to the southern region of the nebula (Fig.~\ref{fig:PSEDfit}) which appears to be freely expanding with little or no interaction with the winds from the nearby, northern WC stars.

We estimate a mass-loss rate of $\sim10^{-3}$ $M_\odot$/yr during the nebula's dust forming phase from Eq.~\ref{eq:MD4}, assuming that $a_\mathrm{max}=60$ $\AA$, $X_g = 0.003$, $v_{t0} = 1$ km/s, $R_f\sim10^{15}$ pc, $v_w=100$ km/s and $\rho_b=3$ g $cm^{-3}$. Although this mass-loss rate implies the duration of the dust production is about the dynamical timescale of the nebula, given a total nebular mass of $\sim$9 $M_\odot$, we note that in our derivation of the grain growth rate we assumed 100$\%$ sticking probabilities for the grain collisions and ignored the possible exhaustion of the condensible species. This mass-loss rate should therefore be treated as a lower limit.

The gradient of decreasing maximum grain size from the south to the north of the Pistol nebula (Fig.~\ref{fig:PSEDfit}e) suggests that the grains are being sputtered by the high velocity $v_w\sim2000$ km/s winds from the nearby WC stars north of the nebula. A grain encountering the WC star winds of velocity, $v_{WC}$, is sputtered at a rate of 

\beq
\frac{d N}{d t}=2 \pi \,a\,v_{WC}\,n_H\,A_i\,Y_i(E_i),
\label{eq:sput1}
\eeq

where $a$ is the grain size, $n_H$ is the hydrogen density of the wind, $A_i$ is the sputtering ion abundance, and $Y_i(E_i)$ is the sputtering yield as a function of sputtering ion's kinetic energy, $E_i$. The rate the grain size decreases can then be given by

\beq
\frac{d a}{d t}=\frac{m_g}{2 \rho _b}v_{WC}\,n_H\,A_i\,Y_i(E_i),
\label{eq:sput2}
\eeq

where $m_g$ is the mass of the sputtered atom and $\rho_b$ is the bulk density of the grain.

With the analytically derived sputtering yields from Tielens et al. (1994) we find that the total change in grain size due to sputtering from the WC winds over $10^4$ yrs is $\sim20$ $\AA$ at the north of the nebula, assuming it is at a distance of $1$ pc from the WC stars (a factor of $\sqrt{2}$ times the projected distance). A more detailed sputtering calculation involving the decelerating expansion velocity of the nebula (Figer et al. 1999) and the decreasing separation between the nebula and the WC stars yields a grain size difference of $\sim15$ $\AA$. 

Our calculated grain size change due to sputtering is consistent with the difference in the maximum grain sizes we observe between the northern and the southern regions of the nebula (Fig.~\ref{fig:PSEDfit}c and e). The preferential sputtering at the north will lead to a deficit of grains in that region, which is what we observe based on the fractional column density around the Pistol Star (Fig.~\ref{fig:PIms}d).

\subsubsection{Pistol Nebula Dust Dynamics}

The geometry of our interpretation of the nebula and WC wind interaction agrees with the observed compression at the north of the nebula. Assuming the WC stars decelerate the projected northern portion of the nebula as it does the back side, we expect a northern expansion velocity of $\sim10$ km/s (Figer et al. 1999). 
From the observed initial and final nebula expansion velocities we can approximate the deceleration of the northern edge and derive the expected distance from the Pistol Star after $10^4$ yrs. The initial expansion velocity, $v_{N0}$, is approximated as the Pistol star wind velocity (100 km/s; Figer et al. 1999 \& Mauerhan et al. 2010) and the final expansion velocity, $v_{N0}'$, is the observed value of $10$ km/s (Figer et al. 1999). We therefore predict the northern edge to be $\sim0.5$ pc from the Pistol Star, which is consistent with the observations.

Using conservation of momentum we derive the following expression for the total mass of the WC winds, $M_{w}$, required to decelerate the northern edge:

\beq
M_{w}=M_{p,N}\left( \frac{v_{N0} - v_{N0}'}{v_w-v_{N0}'}\right)= M_{p,N}\times0.045.
\label{eq:mom1}
\eeq

$M_{p,N}$ is the mass of the northern half of the nebula, and $v_{w}$ is the WC wind velocity. Assuming that $M_{p,N}\sim4.5$ $M_\odot$, half of the observed total nebula mass, the northern edge must encounter $\sim0.2$ $M_\odot$ of particles from the WC winds. The mass flux from the WC winds through the solid angle of the nebula is $\sim0.12\times10^{-4}$ $M_\odot$/yr given a WC mass-loss rate of $10^{-4}$ $M_\odot$/yr; therefore, over $10^4$ yrs we approximate the total WC wind mass swept up by the northern edge of the nebula to be $\sim0.12$ $M_\odot$, which agrees with the $M_{w}$ derived from Eq.~\ref{eq:mom1}.

\subsubsection{The Pistol Nebula vs the LBV3 Nebula}

Unlike the Pistol nebula, the LBV3 nebula freely expands into its surroundings since there are no nearby stars with strong winds to interact with. Due to the lack of near and mid-IR detections of the nebula we are unable to tightly constrain its grain size distribution; however, the distribution of our model fit of the southern Pistol nebula SED ($\sim10 - 60$ $\AA$) lies within the 1-$\sigma$ constraints derived from the DustEM model fits to the LBV nebula SED. The sputtered distribution of the northern Pistol nebula ($10 - 25$ $\AA$) does not fall within the constraints (Fig.~\ref{fig:LBVIms}c). Given the similar nature of the stars themselves (Mauerhan et al. 2010), we may therefore treat the LBV3 nebula as an identical twin to the Pistol nebula with the differences in their appearances only due to the influence of the surrounding environment. Besides the sputtering and compression from external winds, the same evolutionary arguments made for the Pistol can be applied to LBV3.

\subsubsection{Non-detection of a qF362 Nebula}

There are multiple explanations for the non-detection of nebular emission surrounding qF362: the star has already ejected a nebula that is too diffuse to detect, it has undergone an outburst with a mass-loss rate too low to form grains, or it is less evolved than the Pistol and LBV3 and has yet to go through an outburst phase. Although we cannot test all the cases we can rule out and comment on the feasibility of several of them given the history of H-recombination line observations.

We first address the possibility that qF362 may have already formed a nebula that is now too diffuse for detection. From the expression for the optical depth of a dusty shell, $\tau_v=\frac{M\kappa_V}{4\pi v_{exp}^2 t^2}$, where $M$ is the mass of the shell and $\kappa_V$ ($\sim100$ $cm^2$ $g^{-1})$ is the visual opacity of the dust, we can estimate the timescale, $t$, for detectability. Assuming that the nebula has properties similar to the LBV3 and Pistol nebulae ($M\sim6$ $M_\odot$ and $v_{exp}\sim70$ km/s) and that we can detect the nebula until $\tau_V=0.01$, the average optical depth of the Pistol nebula, it should be observable for $\sim1.5\times10^4$ yrs. If indeed the LBV lifetime is $\sim10^4$ yrs we rule out the possibility of qF362 having formed already formed such a nebula.

The recent HST/NICMOS Paschen-$\alpha$ observations (Dong et al. 2011) taken in June 2008 reveal that, unlike the Pistol and LBV3, qF362 is not producing a strong Paschen-$\alpha$ emission line; however, from May 1999 spectral observations of qF362 in the H and K band, Najarro et al. (2009) found that it had similar H-recombination line strengths to that of the Pistol Star. This implies that qF362 is either expanding and cooling or entering an outburst phase of increased mass-loss. 

If it is currently undergoing dust formation we would expect the flux at 8 and 19.7 $\mu$m of the optically thick dust shell surrounding the star to be $\sim170$ and $\sim40$ Jy, respectively, assuming it has a radius of $10^{15}$ cm, the formation radius, and a temperature of $1500$ K. We conclude that dust is unlikely to be forming since a source with a 19.7 flux of 40 Jy would be easily detectable given our integration time and sensitivity. The 8 $\mu$m Spitzer/IRAC only detects a 2.2 Jy source at the location of qF362 (Stolovy et al. 2006) which is more consistent with a star having a luminosity and effective temperature identical to that of the Pistol.

\section{Conclusions}
We have presented imaging of the Pistol and LBV3 nebulae at 19.7, 25.2, 31.5, and 37.1 $\mu$m, tracing the warm dust emission. Our conclusions on the properties and evolution of the LBV3 and Pistol nebulae are summarized in Tab.~\ref{tab:CC}. The analysis suggests that both nebulae formed under very similar stellar conditions; however, the differences in their surrounding environments have differentiated their morphologies. Due to its proximity to the Quintuplet Cluster, the dust in the Pistol nebula is luminous, compressed, and externally heated while the dust in the LBV3 nebula is dim, symmetric, and centrally heated. Interestingly, the gas in the LBV3 nebula is externally ionized by the Arches Cluster and possibly the Quintuplet Cluster, while the gas in the Pistol nebula is externally ionized by the Quintuplet Cluster. Both nebulae share identical size scales, gas-to-dust mass ratios, as well as total gas masses. 

We hypothesize that the Pistol nebula is composed of a population of very small, transiently-heated grains that are preferentially sputtered at the northern region of the nebula where it is colliding with the high-velocity winds from the nearby WC stars. Although the grain composition is not as well determined for the LBV3 nebula as for the Pistol, we find that the non-sputtered southern grain distribution of the Pistol appears to resemble that of LBV3, and thereby supports our interpretation of LBV3 as a twin of the Pistol nebula. 

The non-detection of any emission surrounding qF362 in any the FORCAST wavebands suggests that it is the only one of the three LBVs in the region to not have undergone a dust-forming outburst phase. Any dust that might have been produced by qF362 in sufficient quantities to enshroud the star would have easily been observed by either FORCAST or IRAC.

\emph{Acknowledgments}. We would like to thank the rest of the FORCAST team, George Gull, Justin Schoenwald, Chuck Henderson, and Jason Wang, the USRA Science and Mission Ops teams, and the entire SOFIA staff. This work is based on observations made with the NASA/DLR Stratospheric Observatory for Infrared Astronomy (SOFIA). SOFIA science mission operations are conducted jointly by the Universities Space Research Association, Inc. (USRA), under NASA contract NAS2-97001, and the Deutsches SOFIA Institut (DSI) under DLR contract 50 OK 0901. Financial support for FORCAST was provided by NASA through award 8500-98-014 issued by USRA.

\begin{deluxetable*}{c|ccccccc}
\tablecaption{Summary of Pistol and LBV3 Nebulae Properties}
\tablewidth{0pt}
\tablehead{  & $d$ (pc) & $T_\mathrm{d}$ (K) & $L_\mathrm{IR}$ ($L_\odot$) & $L_*$ ($L_\odot$) &$\tau_{37.1}$ ($10^{-4}$)& $M_\mathrm{dust}$\tablenotemark{a} ($M_\odot$) &  $M_\mathrm{gas}$\tablenotemark{b} ($M_\odot$) }

\startdata
	Pistol&1.2 - 1.4& 100 -180  & $5.2\times10^5$& $3.3\times10^6$ & 3 - 9& 0.03 & 9.3\\ 
	LBV3&1.4& 100&$\sim10^5$& $ 4\times10^6$ &3 - 5& $\sim0.02$ & 6.2\\ 
\enddata

\tablenotetext{a}{Determined by DustEM model fit to observed SED}
\tablenotetext{b}{Derived from free-free/Paschen-$\alpha$ emission}
	\label{tab:PistolLBV3Prop}
\end{deluxetable*}

\begin{deluxetable*}{c|cccc}
\tablecaption{Observed Flux (Jy) from the Pistol and LBV3 nebulae}
\tablewidth{0pt}
\tablehead{ & $F_{19.7}$ & $F_{25.2}$& $F_{31.5}$& $F_{37.1}$}

\startdata
	Pistol& 339 &  585 & 600 & 516 \\ 
	LBV3& 40 & 106 & 158 & 165 \\ 
\enddata

\tablecomments{Flux is given in units of Jy. 3-$\sigma$ error is $20\%$.}
	\label{tab:fluxes}
\end{deluxetable*}

\begin{deluxetable*}{c|cccccccc}
\tablecaption{DustEM Fitting Parameters.}
\tablewidth{0pt}
\tablehead{&$d_*$ (pc)&$d_\mathrm{C}$ (pc)& $L_*$ ($L_\odot$)&$T_*$ (K)&$L_\mathrm{C}$ ($L_\odot$)&$T_\mathrm{C}$ (K)& $a_\mathrm{min}$ ($\AA$)& $a_\mathrm{max}$ ($\AA$)}

\startdata
	   Pistol (PP)&0.5 - 0.7&2.0&3.3 $\times10^6$&12000& 3 $\times10^7$& 35000 & 10 & 25 - 60  \\ 
	   Pistol ($r^{-2}$)&0.5 - 0.7&1.8 - 2.7 &3.3 $\times10^6$&12000& 3 $\times10^7$& 35000 & 10 & 26 - 40  \\ 
               LBV3&0.7&   14   &4 $\times10^6$   &12000&   6 $\times10^7$   &   35000         &       $170^{+430}_{-160}$     & $210^{+1490}_{-170}$ \ 
\enddata

	\label{tab:Pfits}
\end{deluxetable*}

\begin{deluxetable*}{ccccccc}
\tablecaption{LBV3 nebula Intensity Model Parameters}
\tablewidth{0pt}
\tablehead{$r_1$ (pc) & $r_2$ (pc) & $n_0$\tablenotemark{a}  ($cm^{-3}$)& $T_0 (K)$ & $a_\mathrm{min}$ ($\AA$)& $a_\mathrm{max}$ ($\AA$)&$M_d$ ($M_\odot$)}

\startdata
	0.66 & 0.78 & 300 & 95 & 10  & 60 & 0.02\\
\enddata

\tablenotetext{a}{We assume a flat radial density profile}

	\label{tab:LBV3tab}
\end{deluxetable*}
\clearpage

\begin{deluxetable}{ l | p{5.5 cm} p{5.5 cm}}
\tablecaption{Pistol vs LBV3}
\tablewidth{0pt}
\tablehead{ & \colhead{\textbf{Pistol Nebula}} & \colhead{\textbf{LBV3 Nebula}}}

\startdata
	\textbf{Morphology}& $\cdot$ Shell-like and compressed at north and west edge. Shaped by interaction with winds from nearby WC stars.&  $\cdot$ Shell-like and spherically symmetric about star. \\ \\ \hline 
		\textbf{Observed Dust Properties}& $\cdot$ $T_{19/37}$ $\sim 100 - 180$ K decreasing from north to south and $\bar{\tau}_{37}\sim 5 \times10^{-4}$ & $\cdot$ $T_{19/37}\sim 105$ K and $\bar{\tau}_{37}\sim 3.5 \times10^{-4}$ \\ \\ \hline
	\textbf{Ionization} & $\cdot$ Externally ionized by Quintuplet Cluster & $\cdot$ Exernally ionized by Arches Cluster and possibly the Quintuplet Cluster\\ \\ \hline
	\textbf{Model Dust Properties} & $\cdot$ Composed of silicates with sizes 10  $-\,\sim25 -60$ $\AA$ with maximum sizes decreasing from the south to the north. 
	\newline
	\newline
	$\cdot$ $T_d\sim125$ K, $M_d\sim0.03$ $M_\odot$, and $L_\mathrm{IR}\sim5\times10^5$ $L_\odot$
	\newline
	\newline
	$\cdot$ Gas-to-dust mass ratio of  $310^{+77}_{-52}$& $\cdot$ Composed of silicates with minimum grain size $< 600$ $\AA$ and maximum size $40 - 2700$ $\AA$. 	\newline
	\newline
	$\cdot$ $T_d\sim95$ K, $M_d\sim0.02$ $M_\odot$, and $L_\mathrm{IR}\sim10^5$ $L_\odot$ 
	\newline
	\newline
	$\cdot$ Gas-to-dust mass ratio of $293^{+73}_{-101}$\\ \\ \hline
	\textbf{Nebular Evolution}& $\cdot$ Ejected $\sim10^4$ yrs ago with a mass loss rate of $\sim10^{-3}$ $M_\odot$/yr.
	\newline
	\newline 
	$\cdot$ High velocity winds from WC stars decelerate and sputter grains in the northern regions.
		\newline
	\newline
	$\cdot$ Dust heating dominated externally by Quintuplet Cluster  & $\cdot$ Ejected $\sim10^4$ yrs ago with a mass loss rate of $\sim10^{-3}$ $M_\odot$/yr.
		\newline
	\newline
	$\cdot$ Freely expanding
		\newline
	\newline
	$\cdot$ Dust heating dominated centrally by LBV3 \\ 
\enddata

	\label{tab:CC}
\end{deluxetable}

\end{document}